\documentclass[
  journal=pasa,
  manuscript=research-paper,
  year=2024,
  volume=XX,
]{cup-journal}

\usepackage{amsmath}
\usepackage[nopatch]{microtype}
\usepackage{booktabs}

\usepackage{siunitx}
\usepackage{amssymb}
\usepackage{multirow}
\usepackage{subcaption}
\usepackage{hyperref}
\usepackage{float}

\hypersetup{colorlinks=true,citecolor=blue,linkcolor=blue,urlcolor=blue}

\title[Orbital Period Variations of HS 0705+6700]{Orbital Period Variation Analysis of the HS 0705+6700 Post-Common Envelope Binary
}

\author{Huseyin Er}
\affiliation{Atatürk University, Faculty of Science, Department of Astronomy and Space Science, Yakutiye, 25240, Erzurum, Türkiye}
\email[Huseyin Er]{huseyin.er@atauni.edu.tr}

\author{Aykut Özdönmez}
\affiliation{Atatürk University, Faculty of Science, Department of Astronomy and Space Science, Yakutiye, 25240, Erzurum, Türkiye}

\author{Ilham Nasiroglu}
\affiliation{Atatürk University, Faculty of Science, Department of Astronomy and Space Science, Yakutiye, 25240, Erzurum, Türkiye}

\author{Muhammet Emir Kenger}
\affiliation{Atatürk University, Graduate School of Natural and Applied Sciences, Department of Astronomy and Astrophysics, Yakutiye, 25240, Erzurum, Türkiye}

\keywords{binaries: close – binaries: eclipsing -- stars: individual (HS 0705+6700)- subdwarfs -- stars: planetary system} 

\begin{document}

\begin{abstract}

To detect additional bodies in binary systems, we performed a potent approach of orbital period variation analysis. In this work, we present 90 new mid-eclipse times of a short-period eclipsing binary system. Observations were made using two telescopes from 2014 to 2024, extending the time span of the $O-C$ diagram to 24 years. The data obtained in the last seven years indicate significant deviations in the $O-C$ diagram from the models obtained in previous studies. We investigated whether this variation could be explained by mechanisms such as the LTT effect or Applegate. To investigate the cyclic behaviour observed in the system with the LTT effect, we modelled the updated $O-C$ diagram using different models including linear/quadratic terms and additional bodies. The updated $O-C$ diagram is statistically consistent with the most plausible solutions of models that include multiple brown dwarfs close to each other. However, it has been found that the orbit of the system is unstable on short time scales. Using three different theoretical definitions, we have found that the Applegate mechanism cannot explain the variation in the orbital period except for the model containing the fifth body. Therefore, due to the complex nature of the system, further mid-eclipse time is required before any conclusions can be drawn about the existence of additional bodies.

\end{abstract}

\section{Introduction}
\label{sec:1}
Celestial systems consisting of binary stars are a popular topic due to their diverse and complex behavior. The examination of close binary systems, especially those involving subdwarf B stars (sdBs) or white dwarfs, has provided a comprehensive understanding of stellar evolution, mass transfer, and gravitational interactions within such systems. Close binary systems of the HW Vir type consisting of subdwarf B (sdB) stars and low-mass components such as white dwarfs or M dwarfs (dM) typically represent a subset with intriguing properties \citep{2001MNRAS.326.1391M, 2002MNRAS.336..449H,  2003MNRAS.341..669H, 2003MNRAS.338..752M, 2011MNRAS.412..487K, 2014A&A...563A..61S, 2014ASPC..481...13S}. The binary systems known as post-common envelope binary (PCEB) provide an important opportunity for understanding various astrophysical phenomena such as the light travel time (LTT) effect that is demonstrated by variations in eclipse timings within binary star systems \citep{2012A&A...540A...8B, 2012MNRAS.427.2812H, 2013ApJ...766...11A, 2014A&A...566A.128L, 2018haex.bookE..96M}. Thus, LTT effect can be used to investigate the possible presence of additional object(s) orbiting close binary systems.

The possible existence of the additional bodies orbiting around a short-period ($P=2.3$ h) eclipsing PCEB system HS 0705+6700 (V470 Cam, $V=14.7$ mag), consisting of an sdB star and a dM component \citep{2001AGM....18S0912D}, has been extensively studied by many researchers. The possible presence of a third body orbiting HS 0705+6700 has been reported \citep{2009ApJ...695L.163Q, 2010Ap&SS.329..113Q, 2012NewA...17..325C, 2012A&A...540A...8B, 2017OAst...26..134B}. \citet{2013MNRAS.436.1408Q} and \citet{2015JBAA..125..284P} detected a positive increase in the orbital period of the system, suggesting that this increase could be due to the presence of a fourth body. \citet{2020MNRAS.499.3071S} indicated the presence of two circumbinary brown dwarfs orbiting the system with a two-body model containing a quadratic term. However, the proposed model is dynamically unstable over a timescale of $10^3$ years. Finally, \citet{2022MNRAS.513.2478M} investigated the orbital period variation of the system for both one- and two-body models using different data set rather than those commonly used in literature. Although the two-body model remained stable over a timescale of $10^7$ years, statistically, the one-body model yielded better results than the two-body model. \citet{2022MNRAS.514.5725P} concluded that none of the models in the literature are consistent with the most recent $O-C$. 

This study contributes observationally \textbf{to} new mid-eclipse times to a comprehensive investigation of the mechanisms underlying orbital period variations in HS 0705+6700. This constrains the parameters of potential additional objects and improves our understanding of the structure of the system, which is known to be complex, and its orbital stability.

\section{Observations and Data Analysis}
\label{sec:2}

We conducted an observation campaign for HS 0705+6700 system between Nov 2014 and Jan 2024 using the 1 m telescope equipped with a 4k $\times$ 4k SI1100 CCD, the $15\times15$ $\mu$m pixel size at the TÜBİTAK National Observatory (TUG T100, Antalya, Türkiye) and 50 cm telescope equipped with Apogee Alta U230 2K CCD with $24\times24$ $\mu$m pixel size at the Türkiye National Observatories (ATA50, Erzurum, Türkiye). It should be noted that the CCD at the ATA50 telescope was replaced with a CMOS QHY268M Pro I camera (a $3.76\times3.76$ $\mu$m pixel size)  after April 4, 2022. Our observations were performed in white light to obtain optimal counts, with exposure times ranging from 3 to 25 seconds depending on the seeing. A standard process was used to reduce the CCD frames, i.e., bias subtraction, flat fielding, dark subtraction, and cosmic ray correction. The reduced CCD frames were performed the differential aperture photometry using the same method as in \citet{2021MNRAS.507..809E}. Thus, we obtained the 90 new primary eclipse light curves of the HS 0705+6700.

The system was observed by \textit{TESS} in sectors 20, 47 and 60 from Dec 2019 to Jan 2023. For each of the three sectors, the sampling time of an image was between 20 and 120 seconds. To download the photometric images, the Lightkurve package\footnote{https://docs.lightkurve.org/} \citep{2018ascl.soft12013L} was used, which provides the ability to download \textit{TESS} data from the public data archive at Barbara A. Mikulski Archive for Space Telescopes (MAST\footnote{https://mast.stsci.edu/}). From \textit{TESS} photometric images in all sectors, we obtained 1093 primary eclipse light curves.

\begin{figure*}[hbt!]
    \includegraphics[width=0.85\textwidth]{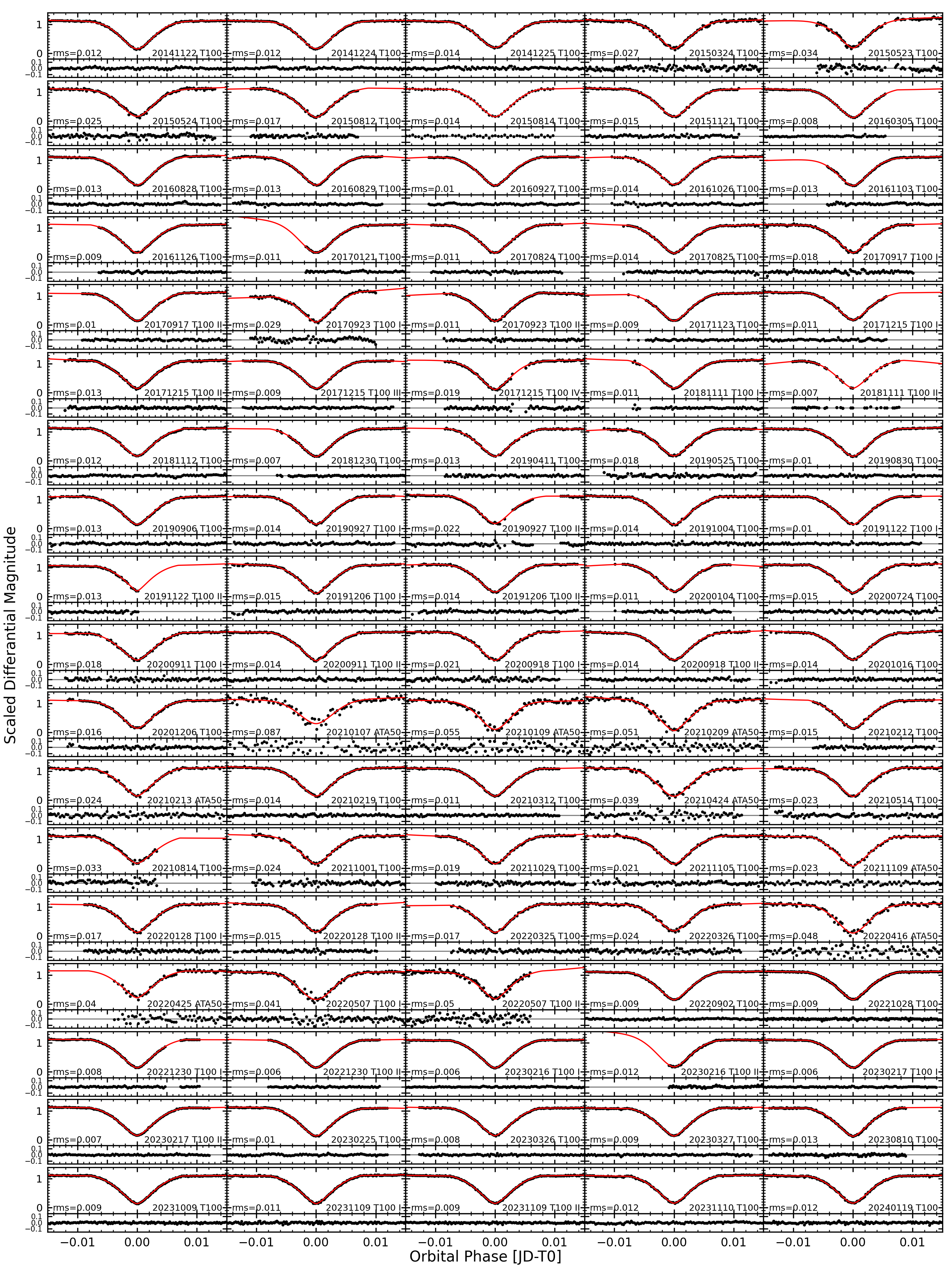}
    \caption{The light curves of HS 0705+6700 during the eclipse were observed using the TUG T100 and ATA50 telescopes. A modified Gaussian profile was used to fit the light curves, as explained in Section \ref{sec:2}. The date of observation and telescope used are labeled, and the sequence of total eclipses on the same day is indicated with Roman numerals. The calculated RMS values, in units of magnitude, are also provided.}
    \label{fig:HS07056700_LC}
\end{figure*}

To determine the mid-eclipse time, we modelled the primary eclipse light curves obtained from our observations and \textit{TESS} data with a modified Gaussian profile as in \citet{2012A&A...540A...8B}. Modelled eclipse light curves obtained from our observations are shown in Figure \ref{fig:HS07056700_LC}. For each of our modeled eclipse light curves, we calculated the root mean square (RMS) from the residuals between observed and modelled light curves. The RMS values range from 0.006 to 0.087 mag with a mean value of 0.017 mag. 

\begin{table}[hbt!]
\begin{threeparttable}
\caption{The mid-eclipse times of HS 0705+6700, its error, references}
\label{tab:midtimes_HS07056700}
\begin{tabular}{lccc}
\toprule
\headrow \multicolumn{3}{c}{In literature} \\
\midrule
BJD & error &  References \\
\midrule
2451822.7605090000 & 0.0000500000 & \citep{2001AGM....18S0912D} \\
2451823.7172080000 & 0.0001000000 & \citep{2001AGM....18S0912D} \\
... & ... & ...\\
\midrule
\headrow  \multicolumn{3}{c}{This work} \\
\midrule
BJD & error &  Telescope \\
\midrule
2456984.5243271800 & 0.0000088821 & T100 \\
2457016.4702787100 & 0.0000091036 & T100 \\
... & ... & ...\\
\midrule
\headrow  \multicolumn{3}{c}{\textit{TESS}} \\
\midrule
BJD & error &  Exposure Time (s) \\
\midrule
2458842.5571969800 & 0.0009745125 & 120  \\
2458842.6529175800 & 0.0034551171 & 120  \\
... & ... & ...\\
\bottomrule
\multicolumn{3}{l}{$^{*}$ Full table is available in its entirety in machine-readable form.}\\
\end{tabular}
\end{threeparttable}
\end{table}

\section{Orbital Period Variations}
\label{sec:3}
We converted all mid-eclipse times to barycentric dynamical Julian time (BJD) using the method described in \citet{2010PASP..122..935E}. Table \ref{tab:midtimes_HS07056700} lists the mid-eclipse times collected from the literature \citep{2001AGM....18S0912D, 2003ASPC..292..129N, 2005IBVS...55..99N, 2007IBVS.5796....1K, 2009ApJ...695L.163Q, 2010Ap&SS.329..113Q, 2012NewA...17..325C, 2012A&A...540A...8B, 2012IBVS.6029....1D, 2013MNRAS.436.1408Q, 2013IBVS.6063....1D, 2015IBVS.6133....1K, 2015IBVS.6153....1P, 2017OAst...26..134B, 2018A&A...611A..48P, 2020JBAA..130..357F, 2020MNRAS.499.3071S, 2022MNRAS.513.2478M, 2022MNRAS.514.5725P}, as well as those obtained from our observations and \textit{TESS} data. Before utilizing this data, we made the subsequent modifications: i) We excluded the outlier mid-eclipse times which scatter more than  three standard deviations from the overall ($O-C$) trend that is calculated every 5000 cycles. ii) Since there are no errors in the times and the start times of the exposures were published for the mid-eclipse times in \citet{2017OAst...26..134B}, we used the mid-eclipse times and errors given in \citet{2020MNRAS.499.3071S} for these mid-eclipse times.  iii)The mid-eclipse times based on data obtained from space-based telescopes are widely scattered in the $O-C$ diagram due to their imprecise nature. Thus, The \textit{TESS} data was binned into five groups based on their cycles, and only these binned times were used during modeling. 

We fit the mid-eclipse times by following  linear ephemeris;

\begin{equation}
\label{eq:lineer_eph}
\begin{aligned}
\ T_{eph}(L) = & T_0 + L \times P_{bin} \\
   = & \text{BJD}\: 2451822.758177(54) + L \times 0.0956467117(8)
\end{aligned}
\end{equation}

In the Equation \ref{eq:lineer_eph}, $T_{eph}$ represents the mid-eclipse time, while $T_0$, $L$ and $P_{bin}$ are the initial ephemeris, the cycle and the orbital period of the binary system, respectively. 

\subsection{Light Travel Time Effect}
\label{sec:3.1}

\begin{figure*}
    \includegraphics[width=0.80\textwidth]{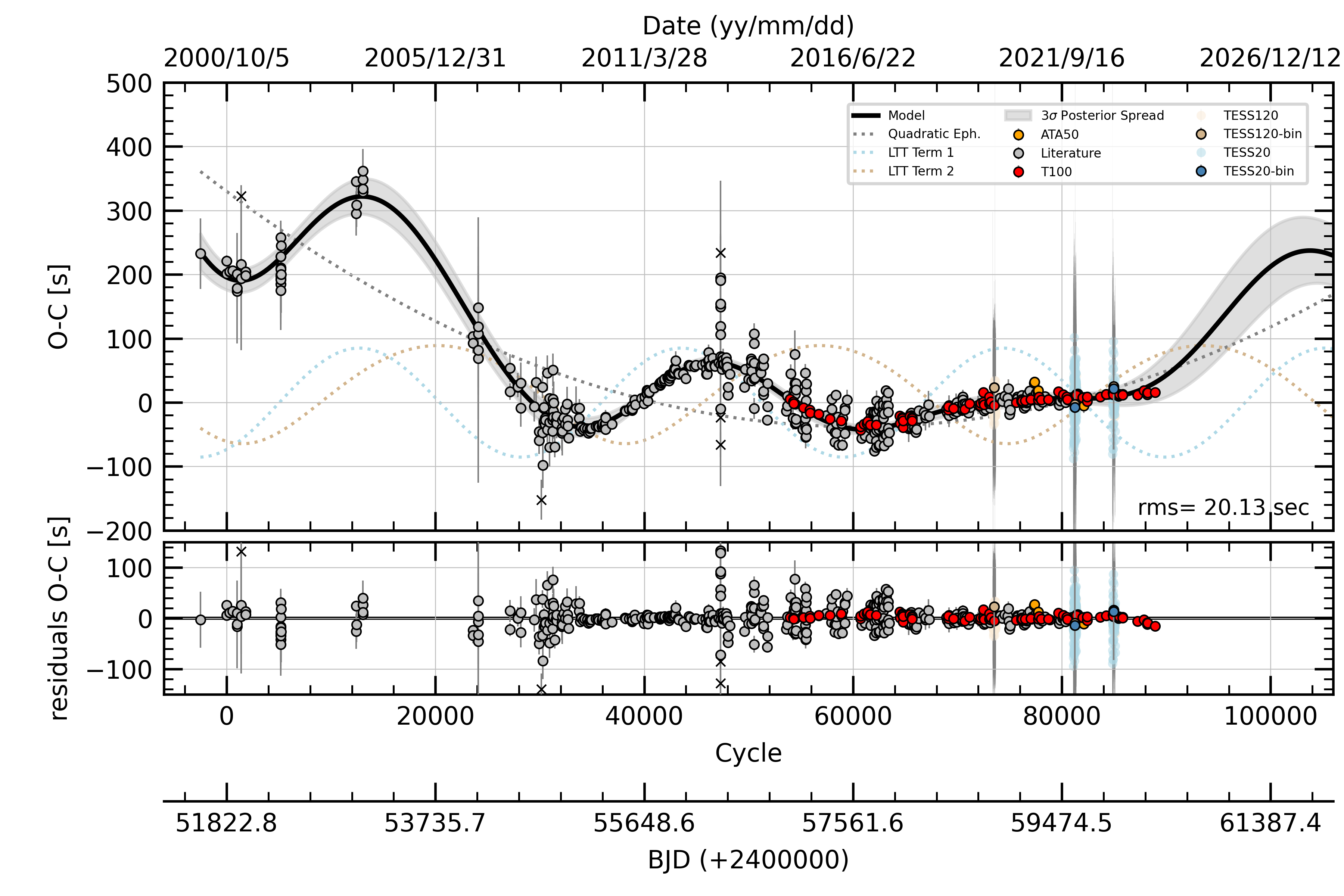}
    \caption{The $O-C$ diagram of HS 0705+6700 is displayed in the upper plot using a linear ephemeris calculated from Equation \ref{eq:lineer_eph}. The TUG T100 is represented by the red-filled circles, while the ATA50 is represented by the orange-filled circles. Data from literature and \textit{TESS} are shown in different colors and labeled with corresponding abbreviations. The model obtained in our study is represented by the black line, which includes the quadratic term and two additional objects. The grey shaded area represents the $\pm3\sigma$ posterior spread. This is calculated from 1000 randomly selected parameter samples from the MCMC posterior. The bottom plot shows the residuals of the $O-C$ times for the model obtained in this study. The figure also includes the calculated RMS value of the residuals of the mid-eclipse times for the model.
    }
    \label{fig:LTT34quad}
\end{figure*}

\begin{figure*}
    \includegraphics[width=0.80\textwidth]{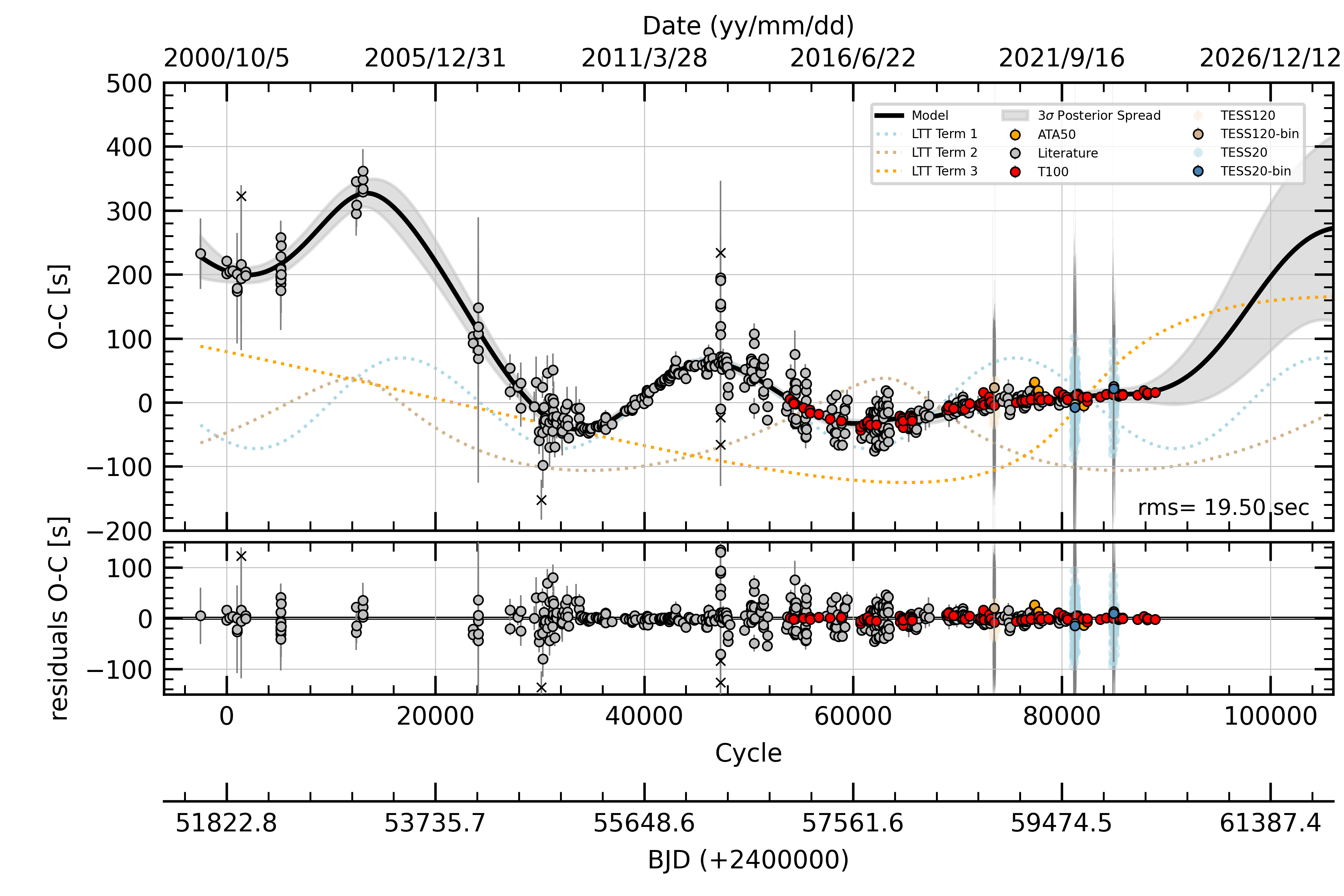}
    \caption{The upper plot displays the $O-C$ diagram of HS 0705+6700 using a linear ephemeris calculated from Equation 
\ref{eq:lineer_eph}.  The model including the three LTT term is represented by the black line. Other informations are as in Figure \ref{fig:LTT34quad}.}
    \label{fig:LTT345}
\end{figure*}

The $O-C$ diagram (see Figure \ref{fig:LTT34quad} and \ref{fig:LTT345}) obtained from residuals of the linear fit indicates a cyclic variation, which can be attributed to the LTT effect. 
To investigate the orbital period variation, we used the models including a quadratic term  ($\beta=P\dot{P}/2$) and/or  the LTT term(s) causing from the presence of hypothetical body. The quadratic term is included to the model by adding $\beta L^2$. The LTT term is defined by \citet{1952ApJ...116..211I}, a modified version provided by \citet{2012MNRAS.425..930G} as following;

\begin{equation}
\label{eq:tau}
 \tau_{i}= K_i \left(\sin{\omega_i} (\cos{E_i (t)} - e_i) + \sqrt{1-e_i^2} \cos{\omega_i} \sin{E_i (t)}\right)
\end{equation}

This formulation is parameterized by Keplerian orbital elements of the N-body companions orbiting around the mass center of the system. In Equation \ref{eq:tau}, $K_i$ is the semi-amplitude of the LTT signal of the $i$ th body, $e_i$ is the eccentricity, $\omega_i$ is the longitude of pericentre, $E_i$ is the eccentric anomaly, $t_{0,i}$ is time of pericentre passage. To prevent weakly constrained values for $e_i$ and $\omega_i$ in quasi-circular and moderately eccentric orbits, we utilized Poincare elements. These elements are represented by $x\equiv e_i cos\omega_i$ and $y\equiv e_i sin\omega_i$  \citep[see][for more details]{2012MNRAS.425..930G, 2015MNRAS.448.1118G, 2017AJ....153..137N, 2023MNRAS.526.4725O}.

This study aims to explain the $O-C$ diagram for each model using various formulations, including those with/without quadratic terms, and with one to three LTT terms.
For example, the model containing a quadratic term and two LTT terms \textbf{are} formulated as follows.

\begin{equation}
 \label{eq:modelquad}
 T_{eph}(L) = T_0 + L \times P_{bin} + \beta L^2 + \tau_{1}(L) + \tau_{2}(L)
\end{equation}

MCMC methodology based on a likelihood function ($\mathcal{L}$) was used to express the orbital period variation, following the identical fitting process described in our previous studies \citep[see][]{2017AJ....153..137N, 2021MNRAS.507..809E, 2023MNRAS.526.4725O}
Uniform prior samples have been randomly assigned to all free parameters within the specified ranges $\beta, K_i, P_i, t_{0,i}, \sigma_f > 0$ days, $x_i,y_i \epsilon$[-0.75,+0.75], $P_{bin} \epsilon$[0.08, 0.15] days and $\Delta T_0 \epsilon$[-10, +10]. The $\mathcal{L}$ function includes the free parameter $\sigma_f$ in units of days to account for systematic uncertainties. This parameter scales the raw uncertainties of eclipsing times ($\sigma_i$) in quadrature. The Markov Chain Monte Carlo (MCMC) method is used to sample the posterior distribution. The sampling process utilized the affine-invariant ensemble sampler implementation from the \textit{emcee} package, following the approach presented by \citet{2010CAMCS...5...65G}, and made available by \citet{2013PASP..125..306F}. For MCMC, 512 initial conditions (walkers) were used to observe the dynamics of each distinct variable in models over 30,000-120,000 steps (depending on used models) within chains. The optimal parameter values, along with their corresponding uncertainties, were determined by assessing the 16th, 50th, and 84th percentiles of the marginalized distributions derived from the maximized likelihood ($L$). The MCMC has been run separately for each models with a variety of formulations.
The models, which include only three LTT terms or one quadratic term and two LTT terms, are the most statistically and astrophysically consistent with the $O-C$ diagram. 
Table \ref{tab:results} presents the most plausible parameters for these models. The best-fitting parameters of the models are shown in the $O-C$ diagram in Figure \ref{fig:LTT34quad} and \ref{fig:LTT345}. Figure \ref{fig:corner_LTT34quad}-\ref{fig:corner_LTT345} consists of the 1D and 2D posterior probability distributions of the system parameters sampled by MCMC. It is important to acknowledge that single-body models aren't consistent with the most recent $O-C$ diagram using all available data. We have discussed the results obtained for all models in Sec \ref{sec:4}.  

The minimum masses of additional bodies can be determined from the following mass function,
\begin{equation}
\label{eq:mass_func}
f(M_i)=\frac{(M_i\sin{i_i})^3}{(M_{i}+M_{bin})^2}=\frac{4\pi^2(a_{12}\sin{i_i})^3}{G P_i^2}
\end{equation}
where $G$ is the gravitational constant, $M_{bin}$ is the total mass of binary, $i_i$ is the inclination of the $i$th body’s orbit, $a_{12}\sin{i_i}$ is the projected semi-major axis of the binary system around the barycentre of the system, $P_i$ is the orbital period, and $M_i$ is the mass of the $i$th body’s. 
We used of $M_{bin}=\:\sim0.617\: M_\odot$ reported by \citet{2001AGM....18S0912D} for the stellar binary mass.

\begin{table}[hbt!]
\begin{threeparttable}
\caption{System Parameters of HS 0705+6700 for two models. }
\label{tab:results}
\begin{tabular}{lll}
\toprule
\headrow Parameters (unit)  & Quadratic + Two Body & Only Three Body \\
\midrule
\multicolumn{3}{c}{For binary} \\
\midrule
$T_0$ (BJD) &  2451822.761996(99) & 2451822.760888(37)\\
$P_0$ (d) &  0.0956465710(34) & 0.0956466921(52)\\
$\beta$ ($10^{-12}$d) & $1.16\pm0.03$ & ...\\
\midrule
\multicolumn{3}{c}{For additional bodies} \\
\midrule
$K_3$ (s)  & $85.17^{+10.94}_{-7.71}$ & $71.04^{+0.45}_{-5.57}$\\
$P_3$ (yr)  & $8.08^{+0.11}_{-0.09}$ & $7.69^{+0.19}_{-0.05}$\\
$x_3$  & $0.0079^{+0.0205}_{-0.0373}$ & $0.0731^{+0.0713}_{-0.0334}$\\
$y_3$  & $0.0032^{+0.0251}_{-0.0368}$ & $0.0193^{+0.0933}_{-0.0244}$\\
$t_{0,3}$ (BJD)  & $2452482^{+1633}_{-3476}$ & $2452873^{+255}_{-147}$\\
$a_{3}\sin{i_3}$ (au)  & $3.48^{+1.15}_{-1.14}$ & $3.36^{+0.05}_{-0.01}$ \\
$e_3$ & $0.0085^{+0.0312}_{-0.0026}$ &  $0.0756^{+0.0528}_{-0.0144}$\\
$\omega_3$ (deg)  &$21.80^{+66.01}_{-109.90}$ & $14.81^{+39.37}_{-19.08}$ \\
$M_3\sin{i_3}$$^*$ ($M_\text{Jup}$)  & $33.30^{+10.68}_{-10.61}$ & $28.57^{+3.66}_{-3.62}$\\
$K_4$ (s)  & $76.69^{+11.02}_{-7.69}$ & $73.68^{+7.55}_{-1.64}$\\
$P_4$ (yr)  & $9.60^{+0.14}_{-0.13}$ & $13.33^{+0.36}_{-0.26}$ \\
$x_4$  &$-0.0316^{+0.0372}_{-0.0311}$ & $-0.2087^{+0.0403}_{-0.1237}$\\
$y_4$  &$-0.1588^{+0.0418}_{-0.0337}$ & $0.4653^{+0.0177}_{-0.1804}$ \\
$t_{0,4}$ (BJD)  & $2451878^{+138}_{-122}$ & $2453126^{+317}_{-74}$ \\
$a_{4}\sin{i_4}$ (au)  & $3.96^{+1.53}_{-1.52}$ &  $4.90^{+0.09}_{-0.07}$\\
$e_4$  & $0.1619^{+0.0998}_{-0.1527}$ & $0.5099^{+0.0288}_{-0.1029}$\\
$\omega_4$ (deg)  & $-101.25^{+182.86}_{-12.09}$ & $114.16^{+23.10}_{-3.60}$\\
$M_4\sin{i_4}$$^*$ ($M_\text{Jup}$)  & $27.45^{+9.92}_{-9.90}$ & $20.96^{+1.91}_{-0.44}$\\

$K_5$ (s)  & ... & $195.82^{+40.31}_{-29.60}$\\
$P_5$ (yr)  & ... & $38.23^{+11.33}_{-6.58}$\\
$x_5$  & ... & $0.6711^{+0.0330}_{-0.1455}$ \\
$y_5$  & ... & $-0.1022^{+0.0687}_{-0.1991}$\\
$t_{0,5}$ (BJD)  & ... & $2445648^{+2358}_{-4257}$\\
$a_{5}\sin{i_5}$ (au)  & ... & $10.03^{+1.91}_{-1.11}$ \\
$e_5$  & ... &  $0.6788^{+0.0443}_{-0.0908}$\\
$\omega_5$ (deg)  & ... & $-8.65^{+5.67}_{-19.49}$\\
$M_5\sin{i_5}$$^*$ ($M_\text{Jup}$)  & ... & $28.34^{+1.33}_{-1.82}$\\

\midrule
\multicolumn{3}{c}{Statistic} \\
\midrule
$\sigma_f$ (s)  & $13.70^{+0.55}_{-0.52}$ & $12.10^{+0.69}_{-0.28}$\\
RMS (s) & 20.13 & 19.50\\

\midrule
\bottomrule
\end{tabular}
\end{threeparttable}
\end{table}

\subsection{Applegate mechanism}
\label{sec:3.3}
The orbital period variation of binary star systems can also be attributed to the magnetic cycle of the low-mass active stars in the system. This is known as the Applegate mechanism \citep{1992ApJ...385..621A}. Changes in the shape of a magnetically active component can contribute to the orbital period variation of the system. 
To test the magnetic mechanism on the $O-C$ signals, we calculated the energy ratios ($\Delta E/E_{sec}$) through three different Applegate approaches, as follows: Thin-shell model \citep{2009Ap&SS.319..119T}, Finite-shell two-zone model \citep{2016A&A...587A..34V}, and Spin–orbit coupling model \citep{2020MNRAS.491.1820L}. 
\begin{table}
\caption{The energy ratios ($\Delta E$/$E_{sec}$) for the formulation of corresponding Applegate mechanisms \citep{2009Ap&SS.319..119T,2016A&A...587A..34V,2020MNRAS.491.1820L}.}
\label{tab:applegate}
\begin{tabular}{l|c|c|c}
\hline
$O-C$ Models & Thin-shell & Finite-shell & Spin–orbit \\
 & model & two-zone model & coupling model \\
\hline
Quadratic+ two LTT   & 6.43 & 121 & 203 \\
Three LTT  & 0.66 & 32.68 & 12.26 \\
\hline
\end{tabular}
\end{table}
For all Applegate models, the magnetic activity could occur under the condition that $\Delta E/E_{sec}$ is less than 1 (i.e. $\Delta E \ll E_{sec}$). For the calculations of the energy ratios, the parameters obtained in Table \ref{tab:results} and determined by \citet{2016A&A...587A..34V} were used. Those calculated for the smaller LTT signal are listed in Table \ref{tab:applegate}. For the LTT signal of the fifth body in the model with three LTT terms, the $\Delta E/E_{sec}$ is calculated to be 0.66 using the thin shell model \citep{2009Ap&SS.319..119T}, depending on solar-like magnetic cycles in the secondary star. The other calculated energy ratios are much higher than the threshold. Thus, only the LTT signal of the fifth body can be attributed to the magnetic cycle in the case of the thin-shell model.

\subsection{Orbit Stability Analysis}
\label{sec:3.2}

To investigate the orbital stability of HS 0705+6700 in our models, we used the N-body orbital integration package of the \textit{REBOUND}\footnote{https://rebound.readthedocs.io} \citep{2012A&A...537A.128R}, which includes a Mean Exponential Growth factor of Nearby Orbits (MEGNO, \citep{2000A&AS..147..205C}) indicator and a Wisdom-Holman symplectic integrator (WHFAST, \citep{2015MNRAS.452..376R}). Using N-body integration, REBOUND simulates the motion of celestial objects and provides two significant insights: First, the MEGNO chaotic parameter surface is mapped, yielding an indicator $<Y>$ that assesses the chaotic behaviour of the system over a range of semi-major axis and eccentricity values over a given period of time.  A stable system is indicated by $<Y>\leq2$, while values greater than 2 indicate chaotic (unstable) orbital configurations. A value of 10 is assigned to $<Y>$ when a particle is ejected or collides. Secondly, the orbital stability timeline integrates the orbits for a given time and shows the variations in parameters such as semi-major axis and eccentricity as a function of time. This is useful for understanding planetary interactions, predicting system escape or collision, and determining the stability period of orbits.

In both simulation scenarios, the central binary star was treated as a singular mass, and all orbital trajectories were confined to a co-planar configuration. We set the optimal timestep for WHFast to be roughly 0.1\% of the shortest orbital period of the additional bodies. It was also assumed that the limit distance for escaping from the system is 20 AU. Dynamic stability simulations were performed using the model parameters to obtain both the MEGNO value and the orbital stability timeline. It has been found that all system configurations constructed from the system parameters of the models in Table \ref{tab:results} are unstable even <2000 yr. In addition, the stability tests were performed under the assumption that the detected signal of the fifth body was raised from the magnetic cycle, yet the stable system configuration on longer time scales can not be constructed.

\section{Discussion and Conclusions}
\label{sec:4}
We present 90 new primary mid-eclipse times for HS 0705+6700 from 2014 to 2024. By combining our mid-eclipse times with those obtained from \textit{TESS} data in this study and from the literature, we analysed the detected orbital period variation in the derived $O-C$ diagram. Our data covers a time span of 10 years, which extends the time span of the $O-C$ diagram by about 2 years, over a total of 24 years.

All possible models were used to test the new $O-C$ diagram. These include models with quadratic/nonquadratic terms and with one to three LTT terms. Studies in the literature explaining the orbital period variation of the HS 0705+6700 with models containing only a single body are based on $O-C$ diagrams prior to 2017. However, \citet{2022MNRAS.513.2478M} reported a single body model consistent with their $O-C$ diagram, without using the eclipse times around 2004 (cycle ~12000) obtained by \citep{2005IBVS...55..99N}. We could not find a valid reason, such as large uncertainty, to exclude all of these times. We couldn't find any plausible model that includes only one LTT term explaining the orbital variability of the most recent $O-C$ diagram using all available data. The statistical coherence of the $O-C$ diagram is maximised when using a model including more than one LTT terms.  Exceptionally, the model including only two LTT signals resulted in very high semi-amplitudes, implying M-type stars with $M_3= 138\: M_\text{Jup}$ and $M_4= 151\: M_\text{Jup}$. It is astrophysically uncommon for a stable quadruple system to contain stars so close together ($a<5$ au). Although the RMS for this model is 21.95s, the posterior probability distributions show bimodality with two solutions of the parameters, so the parameters have high uncertainties. 
Thus, this model is statistically and astrophysically less likely to explain the current $O-C$ diagram.

The system parameters for the two most plausible models are listed in Table \ref{tab:results}. For the first model with a quadratic and two LTT terms, the RMS value is calculated as $20.13$ s. 
For the inner and outer bodies of the model, the semi-amplitudes of the LTT signals were determined to be $K_3=85.17$ and $K_4=76.69$ s, while the orbital periods are $P_3=8.08$ yr and $P_4=9.60$ yr, and the semi-major axis is $a_{3}\sin{i_3}=3.48$ and $a_{4}\sin{i_4}=3.96$ au. These parameters yielded minimum masses of 33.30 and 27.45 $M\text{jup}$, implying brown dwarfs. The quadratic term with a positive coefficient ($\beta=1.16\times10^{-12}$) obtained for this model can be associated with a long-term perturbation caused by an additional body. Thus, we also investigated the $O-C$ diagram with three LTT terms without the quadratic term. The final model provides the best RMS value of $\sim20$ s of all the models. This model includes third brown dwarf with a minimum mass of 28.34 $M_\text{Jup}$ and a semi-major axis of 10.03 au. In the system configuration of this model, the other brown dwarfs have minimum masses of 28.57 $M_\text{Jup}$ and 20.96 $M_\text{Jup}$ (see Table \ref{tab:results}).

The sinusoidal variation in $O-C$ is attributed to the magnetic cycle, as is the case for the LTT effect. The studies in the literature searched for the magnetic cycle through Applegate mechanism \citep[see][]{2016A&A...587A..34V, 2022MNRAS.514.5725P}, but it was reported that the magnetic cycle is not a possible explanation for the orbital period variation of the HS 0705+6700.  
Our investigation of the magnetic cycle for orbital period variation includes the three different modified Applegate models using the parameters for the LTT term with the smallest amplitude (see Table \ref{tab:applegate}). 
In the case of the fifth body in the model including three LTT terms, the energy ratios are calculated close to the required energy limit only for the thin-shell magnetic mechanism. It suggests that the magnetic cycle is potentially responsible for the periodic signal in the $O-C$ caused by the fifth body in the system.
The orbital period variation in the other models cannot be explained by the Applegate mechanisms alone due to much higher energy ratios than the threshold limit. 

Although it is possible to obtain a statistical model explaining the $O-C$ diagram, it is important to ensure that the orbits in the system remain stable for at least a few thousand years. We investigated the stability of the orbital configurations constructed for the model parameters in Table \ref{tab:results}. 
The orbital configurations of all models remain unstable and disrupt the system configuration within 2000 years. This agrees with those reported by \citet{2020MNRAS.499.3071S}. The orbital structure of the HS 0705+6700 system appears to be highly complex, according to these results.

The variation in light travel time resulting from the reflex motion of the centre of mass of a HW Vir binary system can be attributed to the presence of one or more orbiting sub-star objects \citep{2012A&A...543A.138B,2016PASP..128h2001H,2018MNRAS.481.2721B,2021A&A...648A..85E,2021MNRAS.506.2122B}. For instance, the existence of additional objects orbiting HW Vir has been postulated on the basis of analysis of eclipse timing variations \citep{2012A&A...543A.138B,2021A&A...648A..85E}.
It is not sufficient to identify the existence of an additional body in motion within the system with the use of the LTT alone. Consequently, Baycroft employed the catalogue of Hipparcos and Gaia proper motion anomalies to demonstrate the existence of a slight indication of a circumbinary companion orbiting HW Vir \citet{2023MNRAS.526.2241B}. It has been reported by \citep{2023MNRAS.526.2241B} that the eventual publication of the complete Gaia epoch astronomy will be an important method to confirm the existence of possible additional components around HW Vir and similar systems. Furthermore, to understand the evolution of binary systems, it is important to investigate additional bodies in evolved star systems and explore potential formation scenarios. There are studies indicating the existence of brown dwarfs in common post-envelope binaries (PCEBs) \citep{2011AIPC.1331...56P, 2013A&A...549A..95Z, 2015A&A...576A.123S}. It was reported that  additional bodies can form before the common envelope (CE) phase and orbits evolve due to changes in gravitational potential for a pure first-generation scenario, while additional bodies can be formed from material ejected during the CE in the second-generation scenario. 
Additionally, a hybrid scenario consists of a combination of the first and second-generation formation scenarios \citep{2015AN....336..458S}. The formation of the brown dwarf(s) as an additional body(ies) is more likely with this hybrid scenario. It is possible that the brown dwarfs in the HS 0705+6700 system formed before CE and evolved during CE. Furthermore, additional planets may have formed from the second-generation disk. For the latest $O-C$ diagram, the brown dwarfs within the complex and chaotic orbital configuration of HS 0705+6700 are found to be responsible for the observed period variations through the LTT effect. Thus, further observations of this system are needed to ultimately understand the orbital configuration, formation, and evolution of the system.


\begin{acknowledgement}
This work has been supported by The Scientific and Technological Research Council of Turkey (TUBITAK), through project number 114F460 (I.N., H.E.). IN, AO, and HE were supported by the Scientific Research Project Coordination Unit of Ataturk University, Project ID 11159. We would like to thank the team of TUBITAK National Observatory (TUG) for partial support in using the T100 telescope (with project numbers TUG T100-631 and TUG T100-1333). We thank the Türkiye National Observatories and the Atatürk University Astrophysics Research and Application Center (ATASAM) for their support in facilitating our use of the ATA50 telescopes. Funding for the ATA50 telescope and the attached CCD has been provided by Atatürk University (P.No. BAP-2010/40) and Erciyes University (P.No. FBA-11-3283) through Scientific Research Projects Coordination Units (BAP), respectively. This paper includes some of the data collected by the \textit{TESS} mission, which are publicly available from the Barbara A. Mikulski Archive for Space Telescopes (MAST) operated by the Space Telescope Science Institute (STScI). Funding for the \textit{TESS} mission is provided by the NASA Science Mission Directorate.

\textit{Software}: Python packages (ccdproc \citep{2017zndo...1069648C}, Astropy \citep{2013A&A...558A..33A}, Numpy \citep{2020Natur.585..357H}, Matplotlib \citet{2007CSE.....9...90H}, Photutils \citep{2020zndo...4049061B}, LMFIT \citep{2014zndo.....11813N}, REBOUND \citep{2012A&A...537A.128R}, emcee \citep{2013PASP..125..306F}, corner.py \citep{2016JOSS....1...24F}, Applegate calculator: http://theory-starformation-group.cl/applegate.
\end{acknowledgement}



\paragraph{Data Availability Statement}
The data underlying this article are available in the article and in its online supplementary material.

\printendnotes

\bibliography{example}

\appendix
\section{}
\begin{figure*}[hbt!]
	\includegraphics[width=\textwidth]{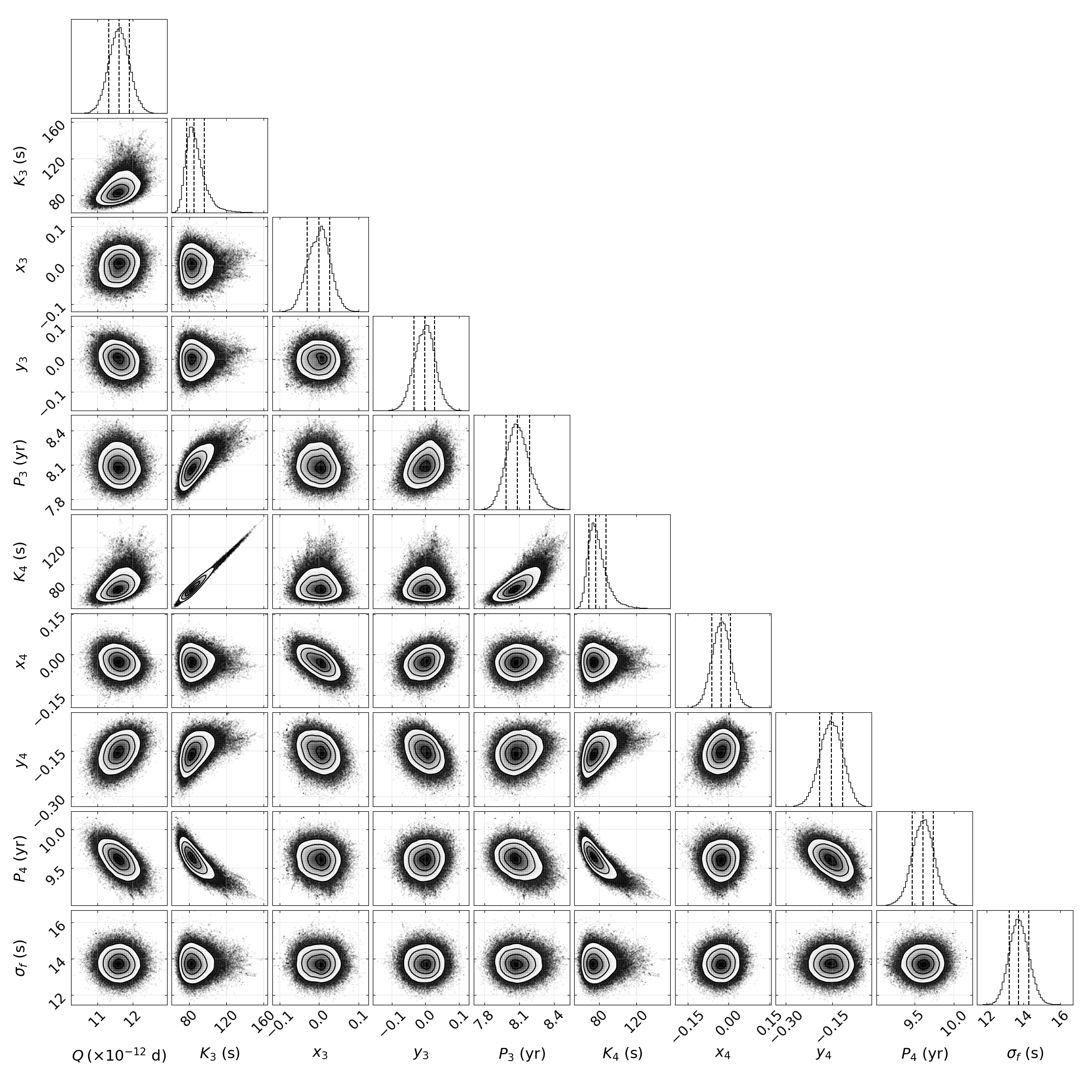}
    \caption{This figure shows the 1-D and 2-D projections of posterior probability distributions of free parameters extracted from the  $O-C$ diagram for the model including quadratic term and two LTT terms , which is made using corner.py \citep{2016JOSS....1...24F}. }
    \label{fig:corner_LTT34quad}
\end{figure*}

\begin{figure*}[hbt!]
	\includegraphics[width=\textwidth]{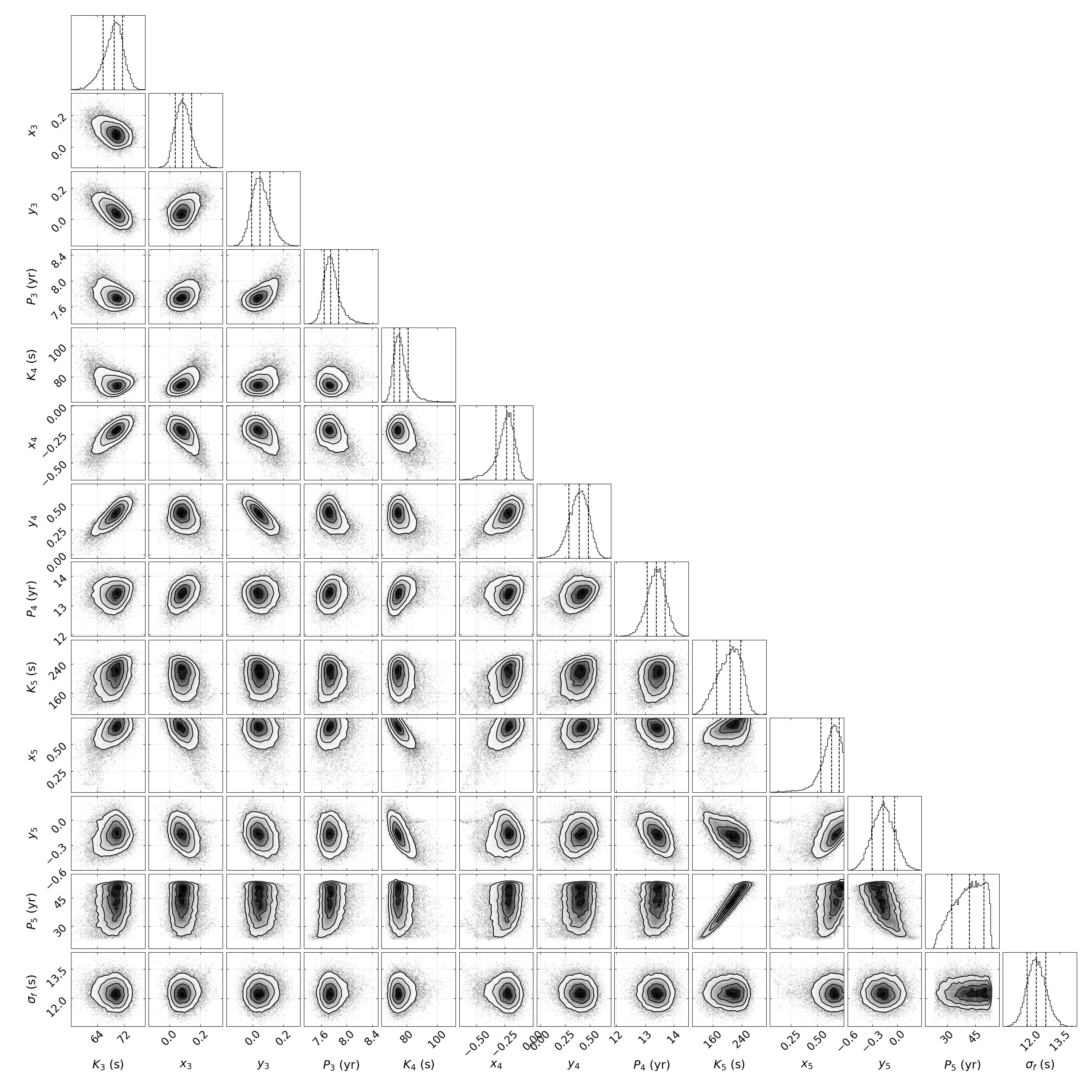}
    \caption{This figure shows the 1-D and 2-D projections of posterior probability distributions of free parameters extracted from the  $O-C$ diagram for the model including three LTT terms, which is made using corner.py \citep{2016JOSS....1...24F}. }
    \label{fig:corner_LTT345}
\end{figure*}

\end{document}